\documentclass[ALICE,manyauthors,onecolumn]{cernphprep}
\usepackage[comma,square,numbers,sort&compress]{natbib}
\usepackage{hyperref}
\usepackage{lineno}
\usepackage{multirow}
\usepackage{amsmath}
\usepackage{siunitx}
\usepackage{xcolor}
\usepackage{wasysym}
\usepackage{placeins}
\usepackage{bm}
\usepackage{comment}
\usepackage[T1]{fontenc}
\usepackage{orcidlink}
\DeclareSIUnit\fm{\femto\meter}
\DeclareSIUnit\eVc{\eV\per\clight}
\DeclareSIUnit\keVc{\keV\per\clight}
\DeclareSIUnit\MeVc{\MeV\per\clight}
\DeclareSIUnit\MeVcc{\MeV\per\clight\squared}
\DeclareSIUnit\GeVc{\GeV\per\clight}
\DeclareSIUnit\GeVcc{\GeV\per\clight\squared}
\DeclareSIUnit\clight{\text{\ensuremath{c}}}
\DeclareSIUnit[number-unit-product = ]\percent{\char`\%}

\newcommand{\onethree}     {$\sqrt{s}~=~13$~Te\kern-.1emV}

\newcommand{\vzero}{$\rm{V}^{0}$\,}

\newcommand{\LaL}{\ensuremath{\mbox{$\Lambda$~($\overline{\Lambda}$)}}}
\newcommand{\XiaXi}{\ensuremath{\mbox{$\Xi^{-}$~($\overline{\Xi}^{+}$)}}}
\newcommand{\XizaXiz}{\ensuremath{\mbox{$\Xi^{0}$~($\overline{\Xi}^{0}$)}}}
\newcommand{\SigmaaSigma}{\ensuremath{\mbox{$\Sigma^{0}$~($\overline{\Sigma}^{0}$)}}}
\newcommand{\OaO}{\ensuremath{\mbox{$\Omega^{-}$~($\overline{\Omega}^{+}$)}}}

\newcommand{\pp}{pp}
\newcommand{\ppNoSpace}{\ensuremath{\mathrm {p\kern-0.05em p}}}

\newcommand{\LXi}{\ensuremath{\mbox{$\Lambda$--$\Xi^{-}$}}}

\newcommand{\aLaXi}{\ensuremath{\mbox{$\overline{\Lambda}$--$\overline{\Xi}^{+}$}}}

\newcommand{\pOm}{\ensuremath{\mbox{p--$\Omega^{-}$}}}

\newcommand{\kstar}{\ensuremath{k^{*}}} 
\newcommand{\XiXi}{\ensuremath{\mbox{$\Xi^{-}$--$\Xi^{-}$}}}

\newcommand{\MeVc}{\ensuremath{\mathrm{MeV}\kern-0.05em/\kern-0.02em \textit{c}}}
\newcommand{\MeVcc}{\ensuremath{\mathrm{MeV}\kern-0.05em/\kern-0.02em \textit{c}^2}}
\newcommand{\GeVc}{\ensuremath{\mathrm{GeV}\kern-0.05em/\kern-0.02em \textit{c}}}
\newcommand{\GeVcSq}{\ensuremath{\mathrm{GeV}\kern-0.05em/\kern-0.02em \textit{c}^2}}
\newcommand{\MeVcSq}{\ensuremath{\mathrm{MeV}\kern-0.05em/\kern-0.02em \textit{c}^2}}

\newcommand{\chEFT}{$\chi$EFT}
\newcommand{\HALQCD}{HAL QCD}

\newcommand{\midrapNpart}{\ensuremath{30}}

\newcommand{\NeVwithcandidates}{\ensuremath{8.57\times10^6}}

\newcommand{\LtrackcleaningPercentage}{\ensuremath{1.9\%}}
\newcommand{\XtrackcleaningPercentage}{\ensuremath{0.6\%}}

\newcommand{\nPairsFinal}{\ensuremath{6142}}

\newcommand{\nPairsInitialLX}{\ensuremath{5.08 \times 10^{6}}}

\newcommand{\nPairsInitialaLaX}{\ensuremath{4.75 \times 10^{6}}}

\newcommand{\radiuswitherror}{$r_\mathrm{eff} = 1.032^{+0.055}_{-0.056} \text{ fm}$}

\newcommand{\purityL}{\ensuremath{95\%}}
\newcommand{\purityX}{\ensuremath{92\%}}

\newcommand{\FractionSelected}{\ensuremath{0.17\%}}

\newcommand{\pzero}{\ensuremath{1.22}}
\newcommand{\pone}{\ensuremath{-8.94 \times 10^{-4}}}
\newcommand{\ptwo}{\ensuremath{8.90 \times 10^{-7}}}

\newcommand{\baselineA}{\ensuremath{0.95}} 
\newcommand{\baselineb}{\ensuremath{2.4 \times 10^{-10}}} 

\newcommand{\genuine}{\ensuremath{32\%}}
\newcommand{\feeddown}{\ensuremath{48\%}}
\newcommand{\misidentification}{\ensuremath{12\%}}
\newcommand{\XiXiresidual}{\ensuremath{8\%}}

\newcolumntype{b}{X}
\newcolumntype{s}{>{\hsize=.5\hsize}X}

\begin{document}%

\begin{titlepage}
\PHyear{2022}
\PHnumber{063}     
\PHdate{24 March}  
\title{First measurement of the $\Lambda$--$\Xi$~interaction in proton--proton collisions at the LHC
}
\ShortTitle{$\Lambda$--$\Xi$~interaction in proton--proton collisions}

\Collaboration{ALICE Collaboration\thanks{See Appendix~\ref{app:collab} for the list of Collaboration members}}
\ShortAuthor{ALICE Collaboration} 

\begin{abstract}
The first experimental information on the strong interaction between $\Lambda$ and $\Xi^-$ strange baryons is presented in this Letter.
The correlation function of $\Lambda$--$\Xi^-$ and \aLaXi\ pairs produced in high-multiplicity proton--proton (pp) collisions at \onethree\ at the LHC is measured as a function of the relative momentum of the pair.
The femtoscopy method is used to calculate the correlation function, which is then compared with theoretical expectations obtained using a meson exchange model, chiral effective field theory, and Lattice QCD calculations close to the physical point.
Data support predictions of small scattering parameters while discarding versions with large ones, thus suggesting a weak \LXi\ interaction.
The limited statistical significance of the data does not yet allow one to constrain the effects of coupled channels like $\Sigma$--$\Xi$ and N--$\Omega$.

\end{abstract}
\end{titlepage}
\setcounter{page}{2}

\section{Introduction}

Understanding the strong interaction among hadrons with strange quarks is one of the main challenges faced by nuclear physics at low energies.
Recent theoretical developments, in parallel with the improvement of computing facilities, enabled
Lattice QCD calculations close to the physical point for systems rich in strangeness~\cite{Sasaki:2019qnh,Ishii:2018ddn,Iritani:2018sra,Gongyo:2017fjb}. 
In contrast to other approaches describing hadron--hadron interactions, such first principles calculations become more stable the higher the (quark) masses involved, and they are expected to deliver reliable results for the interaction of hadrons involving several strange quarks.
On the experimental side, data are scarce for hadron--hadron interactions in the strangeness $|S|>1$ sector due to the difficulties of producing hyperons in large amounts and the fact that those are unstable particles. Hence, the interest in this sector resides in delivering precise data in order to test the ab-initio calculations.

The study of baryon--baryon interactions with strangeness is also crucial for the search of possible di-baryon states beyond the deuteron. 
One debated case is the N$\Omega$ state predicted by Lattice QCD calculations~\cite{Iritani:2018sra} and meson exchange models~\cite{Sekihara:2018tsb}.
It would be held together by an attractive strong interaction at all distances favored by the absence of Pauli blocking in this system. Two-particle correlation studies of \pOm~\cite{STAR:2018uho,ALICE:2020mfd} show that drawing a firm conclusion on its existence is a difficult task due to the complexity of this system that arises from the coupling to several other channels. It has been demonstrated in Ref.~\cite{Kamiya:2019uiw} that the presence of coupled channels modifies the correlation function, in particular in the case of systems size of the order of 1 fm, such as those produced in \pp\ collisions. The thresholds of the $\Lambda$--$\Xi$ and the $\Sigma$--$\Xi$ channels lie around 180 and 90 \MeVcc~below the N--$\Omega$ threshold, respectively; the influence of those channels could severely modify the \pOm~correlation function~\cite{Morita:2016auo,Morita:2019rph} depending on the coupling strength and the characteristics of the $\Lambda$--$\Xi$ and the $\Sigma$--$\Xi$ interactions themselves. The investigation of those channels is thus mandatory in order to clarify the existence of the N$\Omega$ bound state.

Besides the first principles calculations, 
in the sector of nucleon--hyperon (N--Y) and hyperon--hyperon (Y--Y) interactions with strangeness content $|S|\leq3$, several predictions from different theoretical approaches are waiting for validation since over a decade. Leading order (LO) chiral Effective Field Theory (\chEFT)~\cite{Haidenbauer:2009qn}, meson exchange models~\cite{Stoks:1999bz}, and quark constituent models~\cite{Fujiwara:2006yh} produced 
predictions for the $|S|\leq2$ sectors relying on SU(3) symmetry considerations. Those approaches were anchored to the vast database available in the N--N sector, which includes precise determination of scattering cross sections that enable differential studies and partial wave analyses, to measurements of $\Lambda$--N cross sections and, to a lesser extent, to $\Sigma$--N cross sections. 
Very recently, the extension of NLO \chEFT\ potentials from $S=-2$~\cite{Haidenbauer:2015zqb,Haidenbauer:2018gvg} to $S=-3$ and $S=-4$ systems has been explored~\cite{Haidenbauer:2022suy}. These potentials are in accordance with the few experimental constraints on the $\Lambda$--$\Lambda$ and $\Xi$--N interactions and account for effects from SU(3) symmetry breaking in the extension.

At the time when most of those potentials were constructed
there was little hope for
precise data on the interaction between baryon pairs with more than two strange quarks, and the experimental information was limited to the scarce data derived from the detection of hypernuclei, such as binding energies for double-$\Lambda$ hypernuclei
\cite{Danysz:1963zza,Takahashi:2001nm,KEKE176:2009jzw,E373KEK-PS:2013dfg} and $\Xi$ hypernuclei~\cite{Nakazawa:2015joa,J-PARCE07:2020xbm}. However, binding energies alone do not allow unambiguous conclusions on 
the underlying baryon--baryon interactions, and so the available theoretical predictions were awaiting more demanding tests.

Only recently, the femtoscopy technique has delivered precise data with valuable information for the description of the strong interaction among hadrons via the study of two-particle correlations as a function of the relative momentum using collider experiments~\cite{Fabbietti:2020bfg}. Measurements of multi-strange systems $\Lambda$--$\Lambda$ \mbox{\cite{Adamczyk:2014vca,FemtoRun1,FemtoLambdaLambda}}, p--$\Xi^-$~\cite{FemtopXi}, and \pOm~\cite{STAR:2018uho,ALICE:2020mfd} have
been made available from Au--Au collisions at RHIC by STAR and from \pp\ collisions at the LHC by ALICE.

A pioneering study with the first experimental information of the \LXi\ interaction is presented in this Letter. This measurement constitutes the first benchmark for models and theoretical approaches delivering predictions for this system.

\section{Data analysis }
\label{sec:analysis}

The analysis of the correlation function of \LXi\ and \aLaXi\ pairs is performed in the sample of high-multiplicity \pp~collisions at \onethree~collected by ALICE~\cite{Aamodt:2008zz,Abelev:2014ffa} at the LHC during the Run 2 period.
The V0 detector of the ALICE apparatus is used for event selection and triggering. 
The V0 detector consists of two plastic scintillator arrays located on both sides of the collision vertex at pseudorapidities $2.8<\eta<5.1$ and $-3.7<\eta<-1.7$~\cite{Abbas:2013taa}. 
In order to guarantee an uniform acceptance at midrapidity, events are accepted for analysis if the reconstructed primary interaction vertex position along the beam axis ($V_z$) is located no more than 10~cm away from the nominal interaction point. 
In addition, 
a V0-based high-multiplicity trigger is used for selecting events in which the detected signal amplitude exceeds a defined threshold corresponding to the \FractionSelected~highest-multiplicity events out of all inelastic collisions with at least one measured charged particle within $|\eta|<1$ (INEL> 0).
The resulting average charged-particle multiplicity density at midrapidity $(|\eta|<0.5)$ is about $\langle \mathrm{d}N_\mathrm{ch}/\mathrm{d}\eta\rangle =$ \midrapNpart.
An enhanced production of hyperons was reported for such high-multiplicity events~\cite{ALICE:2017jyt}, which provides an abundant sample of hyperons for this analysis.

Strange baryon reconstruction is performed using topological properties of their weak decays, which 
in turn employs the tracking and particle identification capabilities 
of the ALICE detector. 
The Inner Tracking System (ITS) and the Time-Projection Chamber (TPC) are used for charged-particle tracking and momentum reconstruction, the TPC is used as well for particle identification (PID), and the Time-of-Flight detector (TOF) is used for timing information.
The ITS, TPC, and TOF are immersed in a uniform magnetic field along the beam direction with a strength of $\SI{0.5}{T}$. 
The ITS~\cite{Aamodt:2010aa} is made up of six layers of high position-resolution silicon detectors placed at a radial distance between $3.9$ and $\SI{43}{cm}$ around the beam pipe.
The TPC~\cite{Alme:2010ke} is a $\SI{5}{m}$ long gaseous cylindrical detector covering the whole azimuth within the pseudorapidity range $|\eta| < 0.9$. It performs PID by measuring the specific energy loss 
(d$E$/d$x$).
The TOF~\cite{Akindinov:2013tea} consists of Multigap Resistive Plate Chambers which cover the full azimuth range at $|\eta|<0.9$. 
Combinatorial background from out-of-bunch collision pile-up in the TPC is suppressed by rejecting charged tracks unless a matched hit in the ITS, which does not have out-of-bunch pile-up, or in the TOF, with timing information, is present.

The \LaL~are identified exploiting their characteristic V-shaped weak decay $\Lambda \rightarrow \text{p} + \pi^-$ and \mbox{$\overline{\Lambda} \rightarrow \overline{\text{p}} + \pi^+$}, henceforth denoted ``$\rm{V}^{0}$'' decays, and selected within an invariant mass window of 4 \MeVcc\ around the $\Lambda$ nominal mass. Proton and pion tracks are reconstructed using the TPC, and they are identified through the TPC d$E$/d$x$ measurement. They are further combined into \vzero candidates if a certain pair of proton and pion tracks passes a set of geometrical criteria that ensures their consistency with the desired
decay topology. These include a selection on the distance of closest approach (DCA) of the two tracks, the cosine of the pointing angle (CPA) between the line connecting the primary vertex with the candidate's decay vertex and its momentum, and the decay radius.
Furthermore, a minimal transverse momentum of $p_\mathrm{T} \ge 0.3$ \MeVc~is required for the \LaL~candidates.

The \XiaXi~candidates are identified via the weak decay channel $\Xi^{-} \rightarrow \Lambda + \pi^{-}$ and $\overline{\Xi}^{+} \rightarrow \overline{\Lambda} + \pi^{+}$ 
by combining \vzero candidates with a third track with a TPC energy loss signature that is consistent with 
the pion mass hypothesis. They are selected within an invariant mass window of 5 \MeVcc\ around the nominal $\Xi^{-}$ mass. Also in this case, geometric selections serve to identify the expected trajectory 
arrangement and include standard $\Lambda$ and $\overline{\Lambda}$ selections as well as a DCA between the \vzero and the third track and a CPA of the $\Xi$ candidate momentum with respect to the estimated decay position. 
Also the \XiaXi~candidates are required to have a minimal transverse momentum of $p_\mathrm{T} \ge 0.3$ \MeVc.
Candidates are rejected if they are compatible within 5 \MeVcc~with the $\Omega^-$ ($\overline{\Omega}^+$) baryon invariant mass through the weak decay $\Omega^{-} \rightarrow \Lambda + $K$^{-}$ ($\overline{\Omega}^{+} \rightarrow \overline{\Lambda} + $K$^{+}$) by assuming the kaon mass hypothesis for the third track.

The numerical values of the geometrical variables used to select \LaL\ and \XiaXi, as well as the variations of such values used to estimate the systematic uncertainties of the data, are based on previous analyses; see Ref.~\cite{FemtoRun1} and Ref.~\cite{Aamodt:2011zza} for \LaL\ and \XiaXi, respectively.
A fit to the invariant mass spectrum of \LaL~and \XiaXi~ candidates is performed using a double Gaussian to describe the signal and a first-order polynomial for the combinatorial background. For the calculation of the invariant mass of \LaL~the pion and proton hypothesis are used for the daughter tracks; in the case of \XiaXi~ the \LaL~ mass is used for the \vzero and the pion mass for the charged track. The results of the fit deliver average mass resolutions of around 1.5 and 2 MeV/$c^2$ and purities of \purityL~and \purityX~ for \LaL~ and \XiaXi, respectively.

Events are kept for further analysis if at least one \LaL\ and one \XiaXi\ candidate are reconstructed, which results in a total number of \NeVwithcandidates\ events and \nPairsInitialLX\ (\nPairsInitialaLaX) \LXi\ (\aLaXi) pairs.

During the reconstruction, a charged track can be assigned as a decay product of multiple $\rm{V}^{0}$ candidates. For such cases, an
additional selection procedure is implemented to reduce the combinatorial background and choose the best candidate in an unbiased way. The $\rm{V}^{0}$ candidates that share a charged track are discriminated by using several kinematic variables simultaneously, namely the invariant mass, the DCA, and the CPA. They are compared to template distributions from Monte Carlo (MC) simulations and normalized by their expected resolution such that they have the same incidence in the comparison.
For the MC templates, \pp~events are generated using PYTHIA~8.2~\cite{SJOSTRAND2015159} and the resulting particles are propagated through the simulation of the ALICE detector using GEANT3~\cite{Brun:118715}. An analogous track cleaning procedure is used as well for the $\Xi$ candidates that share charged tracks. The track cleaning procedure reduces the sample of $\Xi^-$ and $\overline{\Xi}^{+}$ candidates by 
\XtrackcleaningPercentage~and the sample of $\Lambda$ and $\overline{\Lambda}$ candidates by \LtrackcleaningPercentage.

In order to avoid autocorrelations generated by the pairing of \LaL\ candidates with their mother particles, pairs of \LXi\ (\aLaXi) are not considered if the \LaL\ candidate shares any charged track with the \XiaXi\ candidate.

\section{Correlation function}
\label{sec:CFeval}

Experimentally, the two-particle correlation function is defined as~\cite{Lisa:2005dd}

\begin{equation}
C\left(k^{*}\right)=\mathcal{N} \times \frac{N_{\mathrm{same}}\left(k^{*}\right)}{N_{\mathrm{mixed}}(k^{*})}~,
\end{equation}

where \kstar\ is the relative momentum in the pair rest frame, defined as $k^{*}=\frac{1}{2} \times\left|\mathbf{p}_{1}^{*}-\mathbf{p}_{2}^{*}\right|$ with $\mathbf{p}_\mathrm{i}$ being the three-momenta of the involved particle candidates. 
The $N_{\mathrm{same}}$ is the \kstar\ distribution of particle pairs produced in the same collision, whereas $N_{\mathrm{mixed}}$ is the \kstar\ distribution obtained by pairing particles produced in different collisions with similar $V_z$ and multiplicity.
Due to the event mixing procedure, the number of pairs in $N_{\mathrm{mixed}}$ is higher than in $N_{\mathrm{same}}$, hence the correlation function has to be normalized at large $k^*$, where the effects of the final state interactions are absent. This is denoted by the factor $\mathcal{N}$; the normalization is performed in the region $k^{*} \in[450,650]$ \MeVc. 
In the following, \LXi\ is used to refer to the sum of \LXi\ and \aLaXi\ pairs, since both experience the same interaction and correlation and no significant differences are observed between both pairs during the analysis.
A total of \nPairsFinal\ \LXi\ pairs are found with \kstar\ below 200 \MeVc.

On the theoretical side, the correlation function can be expressed as a function of the particle emitting source $S\left(r^{*}\right)$ and the two-particle pair wave function $\left|\Psi\left(r^{*},k^{*}\right)\right|$ which contains the interaction component~\cite{Lisa:2005dd}
\begin{equation}
C\left(k^{*}\right)=\int \mathrm{d}^{3} r^{*} S\left(r^{*}\right)\left|\Psi\left(r^{*},k^{*}\right)\right|^{2},
\label{eq:CFtheo}
\end{equation}
where $r^*$ refers to the relative distance between the two particles.
The source function is assumed to have a Gaussian shape. Its size is obtained from the universal baryon--baryon transverse mass ($m_\mathrm{T}$) scaling observed in high-multiplicity pp collisions~\cite{Acharya:2020dfb}, and considering its enlargement due to shortly lived resonances~\cite{Acharya:2020dfb}. The average transverse mass of \LXi\ pairs with \kstar$<200$ \MeVc~is $\langle m_\mathrm{T} \rangle  = \SI{2.01}{GeV/\textit{c}^2}$, which leads to an effective source radius of \radiuswitherror. The quoted uncertainties take into account the statistical and systematic uncertainties of the parametrization of the $m_\mathrm{T}$ dependence~\cite{Acharya:2020dfb}.

Given the source size, the theoretical correlation function is computed using two different methods. The Lednický--Lyuboshits (LL) approach~\cite{Ledn,Stavinskiy:2007wb} allows one to calculate the theoretical correlation function when the effective range parameters are known: scattering length $f^s_0$ and effective range $d^s_0$, with $s$ denoting the spin state of the pair. 
Note that in this Letter the standard notation and sign convention in femtoscopy is used, where a positive $f_0$ corresponds to an attractive interaction,
while a negative scattering length corresponds either to a repulsive potential or a bound
state.
For \LXi\ pairs, there are two spin configurations, namely a singlet with $s=0$ and a triplet with $s=1$. They contribute with a weight of $1/4$ and $3/4$ to the total theoretical correlation function, respectively.
The LL model is used to evaluate the predictions from the Nijmegen meson exchange model~\cite{Stoks:1999bz} and the interactions from \chEFT~\cite{Haidenbauer:2009qn,Haidenbauer:2022suy}. The second method uses the CATS framework~\cite{Mihaylov:2018rva}, a Schrödinger equation solver, to evaluate the wave functions for the potentials extracted from Lattice QCD calculations performed by the \HALQCD\ Collaboration~\cite{Ishii:2018ddn}. 

\begin{figure*}[ht]
  \centering
   \includegraphics[width=0.49\linewidth]{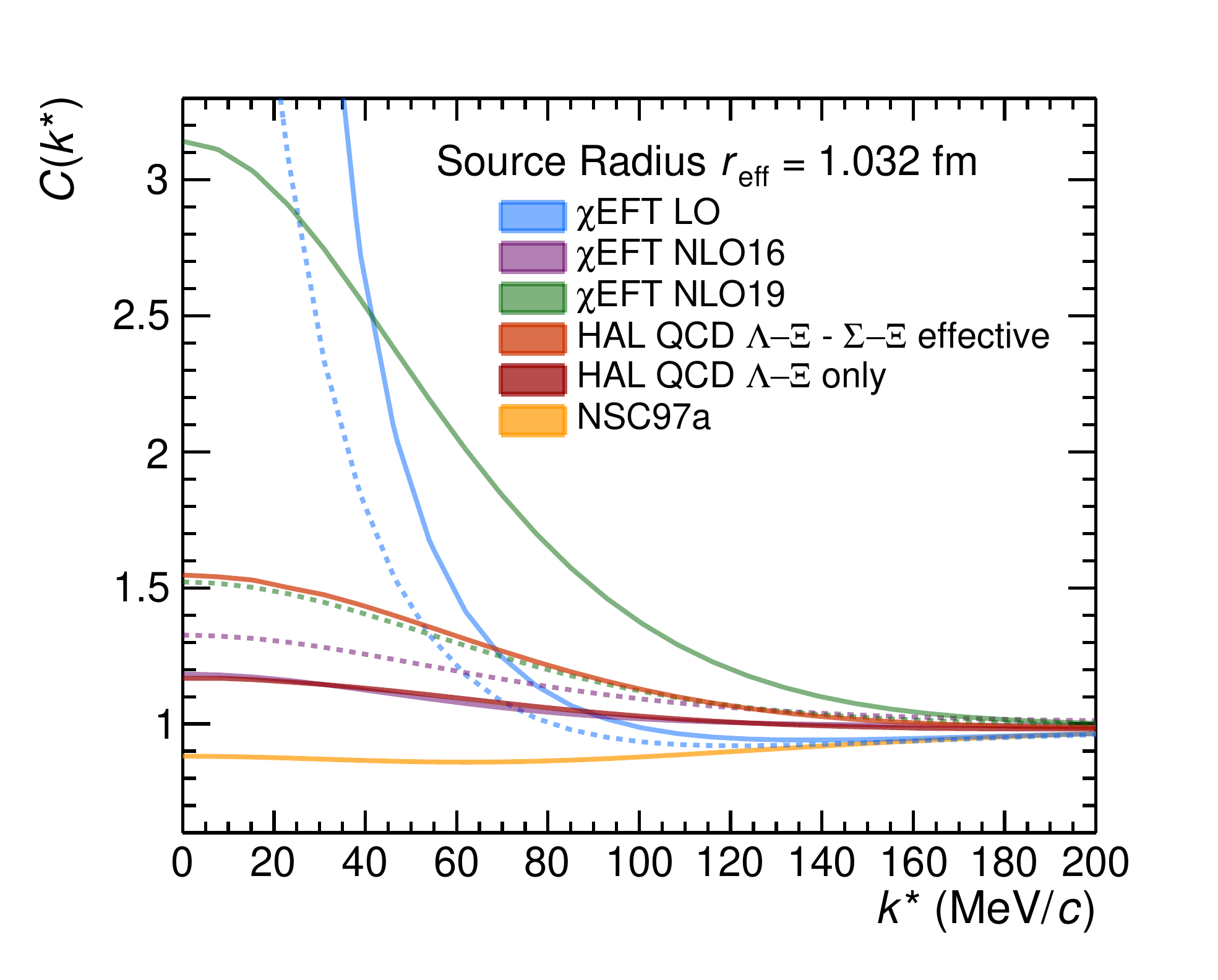}
  \hfill
\caption{Theoretical \LXi\ correlation functions from predictions (\cite{Haidenbauer:2009qn,Haidenbauer:2022suy,Stoks:1999bz,Ishii:2018ddn}) evaluated for the experimental source radius. See text for details.
In the case of the LO and NLO $\chi$EFT potentials the solid (dotted) lines correspond to the lowest (highest) cutoff~\cite{Haidenbauer:2009qn,Haidenbauer:2022suy}.
}
 \label{fig:genuine}
\end{figure*}

The \LXi\ correlation functions evaluated for the central value of the radius $r_\mathrm{eff}$ for each considered theoretical prediction are shown in Fig.\ref{fig:genuine}. Since the Coulomb interaction is absent, the deviations from unity at small relative momentum are exclusively due to the strong interaction.
Details on the characteristics of each interaction are discussed in Section~\ref{sec:Results}.

The experimental correlation function contains additional contributions to the genuine \LXi\ strong interaction as it is defined in Eq.~\ref{eq:CFtheo}. In order to compare the theoretical expectations with the experimental data, a model is built for each theoretical prediction containing all contributions to the experimental correlation function as detected by ALICE
\begin{equation}
C_\textrm{model} \left(k^{*}\right) = C_{\text {non-femto }}(k^*) \times \left[\sum_{i} \lambda_{i} \times C_{i}\left(k^{*}\right) \right],
\label{eq:model}
\end{equation}

where the sum contains all femtoscopic contributions $C_i(k^*)$ namely the genuine as well as contamination induced by the misidentification background and feed-down. Each of them is multiplied with its relative contribution $\lambda_i$.
The $C_{\text {non-femto }}(k^*)$ describes non-femtoscopic effects such as energy conservation which are dominant for large $k^*$. It is modelled phenomenologically by a polynomial with a constant and a third degree term $C_{\textrm{non-femto }}(k^*) = a \times ( 1 - b \times k^{* 3})$, which ensures a flat behaviour at $k^*\rightarrow 0$, and is fitted to the measured correlation function.

In the feed-down contributions, at least one particle of the pair originates from a
decay and appears on timescales larger than the strong interaction which is measured here. They carry the residual correlation from their mother particle which is washed out because of the decay. The main feed-down to the \LaL\ candidates comes from the \XiaXi, \XizaXiz, and \SigmaaSigma\ decays. The combined relative contribution is obtained by fitting the CPA distribution to MC templates. Then, the $\lambda$ parameter for each contribution is determined via isospin considerations. The feed-down to the \XiaXi\ candidates comes from the decays of the resonances $\Xi^{-}(1530)$ and $\Xi^{0}(1530)$ as well as the \OaO\ hyperons. Their $\lambda$ parameters are extracted from the production rates reported in Refs.~\cite{Xi1530,ALICE:2017jyt} and their branching ratios~\cite{10.1093/ptep/ptaa104}.
All feed-down contributions to the correlation function are assumed to be constant in \kstar\ with a value equal to unity, except for the case of the $\Xi^--\Xi^-$ feed-down, with the identified $\Lambda$ coming from the decay of an unidentified $\Xi^-$. The latter contributes with $\lambda_{\XiXi} =$~\XiXiresidual;
it is modelled assuming a pure Coulomb $\Xi^--\Xi^-$ interaction and it is propagated to $C_\textrm{model}(k^*)$ via a momentum transformation from the $\Xi^--\Xi^-$ to the \LXi\ pair rest frame.
The relative contribution of all the other feed-down contributions is of $\lambda_\textrm{flat} =$~\feeddown.

The relative contribution from misidentification is of $\lambda_\textrm{mis.} =$~\misidentification~and is calculated from the purities in the selection of the $\Lambda$ and $\Xi^-$.
This contribution is modelled by a second order polynomial $C_{\textrm{mis.}}(k^*) = p_0 + p_1 k^* + p_2 k^{*2}$ with parameters obtained via a fit to the correlation function constructed using \LXi\ pairs from an invariant mass sideband analysis~\cite{ALICE:2019buq}. The values of the parameters are \mbox{$p_0$~$=$~\pzero}, \mbox{$p_1$~$=$~\pone}~$(\MeVc)^{-1}$, and \mbox{$p_2$~$=$~\ptwo~$(\MeVc)^{-2}$}.

The relative contribution from the genuine \LXi\ interaction is $\lambda_\textrm{genuine} =$~\genuine. In order to consider the finite momentum resolution, evaluated via full simulations of the ALICE apparatus and its response, $C_\textrm{model}\left(k^{*}\right)$ has to be transformed into the basis of the reconstructed momenta as it was done in previous analysis~\cite{FemtoRun1}.

\section{Results}
\label{sec:Results}

\begin{figure*}[ht]
  \centering
   \includegraphics[width=0.49\linewidth]{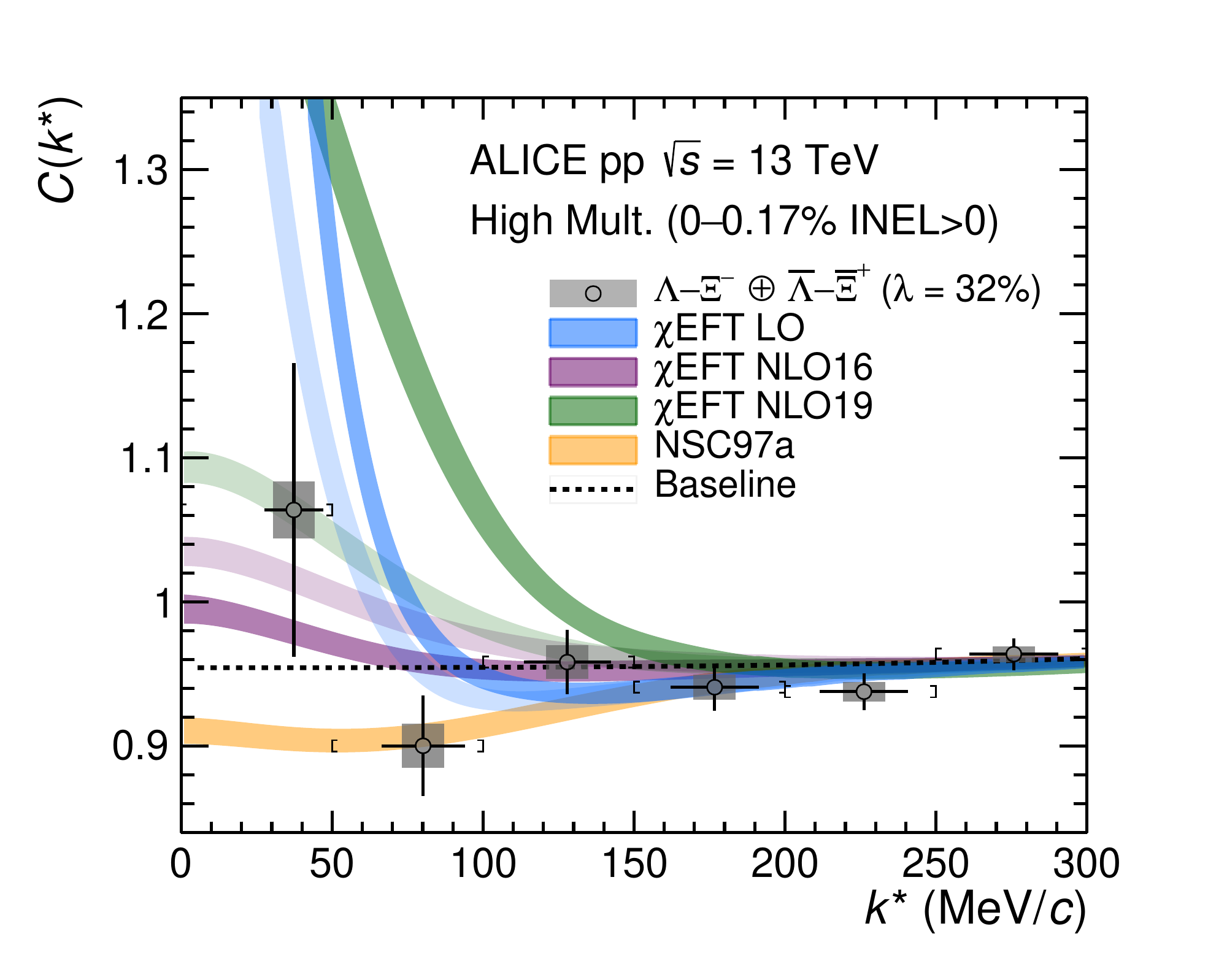}
  \hfill
     \includegraphics[width=0.49\linewidth]{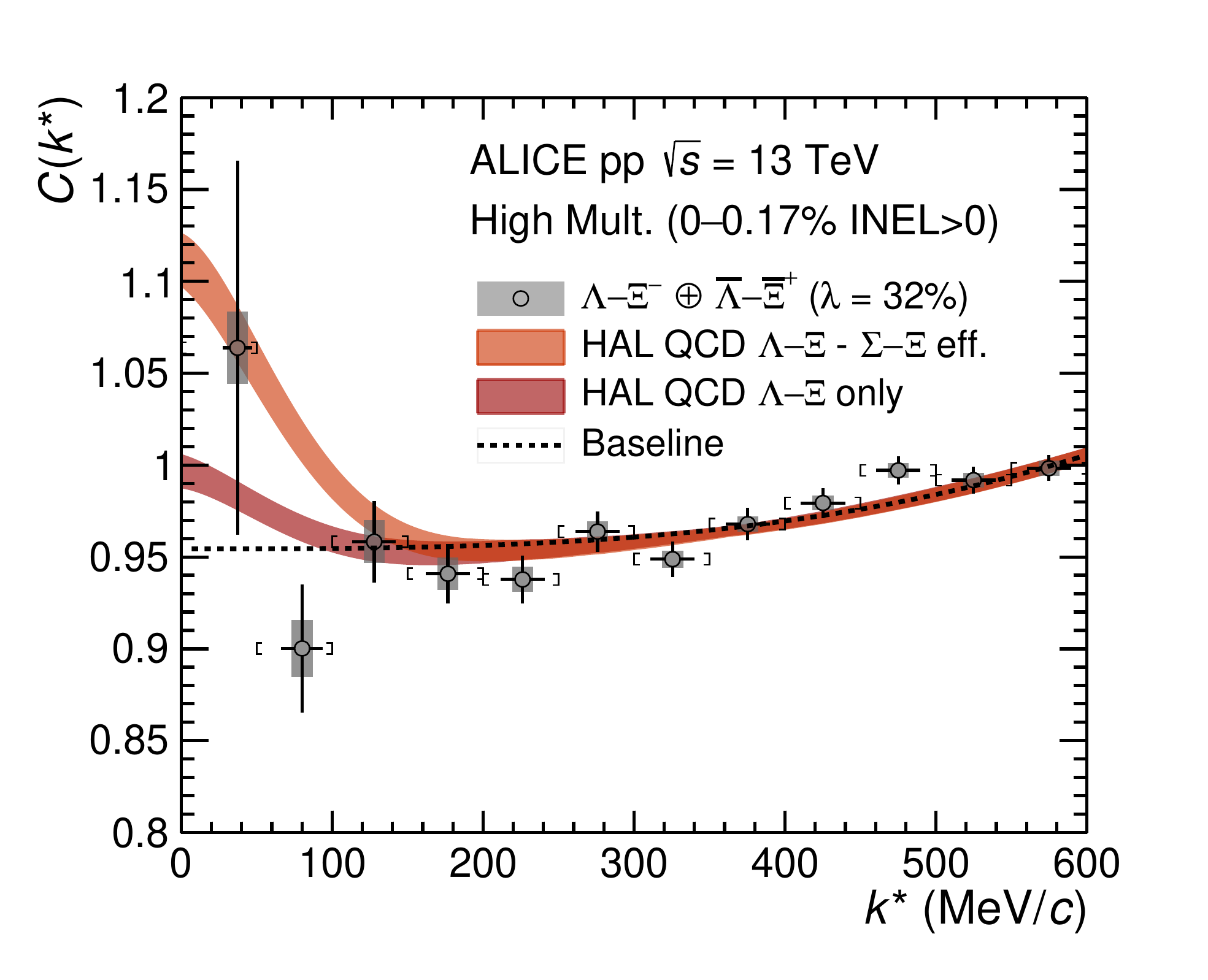}
  \hfill
\caption{Experimental \LXi\ correlation function with statistical (vertical black lines) and systematic (gray boxes) uncertainties. The square brackets show the bin width of the measurement and the horizontal black lines represent the statistical uncertainty in the determination of the mean \kstar\ for each bin. Left panel: Comparison to LO~\cite{Haidenbauer:2009qn} and NLO~\cite{Haidenbauer:2022suy}  \chEFT\ and NSC97a~\cite{Stoks:1999bz} potentials evaluated with the Lednický--Lyuboshits model~\cite{Ledn,Stavinskiy:2007wb}. In the $\chi$EFT models the darker and light bands correspond to the version with the lower and higher cut-off value, respectively. Right panel: Comparison with Lattice QCD calculations by the HAL QCD Collaboration~\cite{Ishii:2018ddn} using an effective potential including the coupling to $\Sigma$--$\Xi$ (orange) and the \LXi\ elastic potential alone (red). The width of the bands in both panels correspond to the systematic uncertainties of the fit as described in the text. The dotted black line represents the result of the baseline fit assuming no \LXi\ strong interaction.
}
  \label{fig:experimentalCF}
\end{figure*}

The experimental \LXi\ correlation function is shown in Fig.~\ref{fig:experimentalCF} in two different \kstar\ and $C(k^*)$ ranges. 
The systematic uncertainties of the data displayed in Fig.~\ref{fig:experimentalCF} are associated with variations on the selection criteria of \LaL\ and \XiaXi\ as explained in Section~\ref{sec:analysis}.
The analysis is repeated with 39 random combinations of such variations. For each \kstar\ point, the final systematic uncertainty is given by the width of a Gaussian fit including all 39 measurements of the correlation function.

The data are compared in Fig.~\ref{fig:experimentalCF} with predictions of the correlation function, 
according to Eq.~\ref{eq:model}, from several theoretical descriptions of the \LXi\ strong interaction and from the assumption of no interaction. The parameter of the non-femtoscopic baseline is fitted to the data for each case in the range \kstar $\in [0,800]$ \MeVc.
The width of the theoretical bands reflect the uncertainties in the evaluation of the correlation function, namely: i) variations of the radius of the source function according to the experimental determination \radiuswitherror; ii) variation of the normalization range by \SI{\pm50}{MeV/\textit{c}}; iii) variation of the range of the fit to the non-femtoscopic baseline by \SI{\pm50}{MeV/\textit{c}}; iv) variation in the parametrization of the baseline using a second order polynomial (i.e. the non-femto contribution becomes $C_{\textrm{non-femto }}(k^*) = a' \times (1 - b' \times k^{* 2})$ ); and v) variation of the functional form describing the sidebands correlation function to $p_0 + \exp (p_1 + p_2 k^*)$.
The evaluation of the theoretical correlation function and the fit of the baseline parameters were performed with all possible combinations of such variations. The width of the theoretical band for each model is given in each \kstar\ point by the root mean squared of all fit results.

The dotted black line in Fig.~\ref{fig:experimentalCF} represents the result of the baseline fit assuming no \LXi\ strong interaction, for which $a=$~\baselineA\ and $b=$~\baselineb~$(\MeVc)^{-3}$ are obtained.
The compatibility with the data is evaluated in terms of the number of standard deviations $n_\sigma$, which were obtained from the p-value computed in the range \kstar$<200$~\MeVc. The uncertainties of the data were considered by adding the statistical and systematic uncertainties in quadrature. The result for the "no strong interaction" assumption is $n_\sigma =$~0.78 showing that in the low relative momentum region, where femtoscopic effects are expected, data do not deviate significantly from the baseline. 

The dark blue and light blue bands in the left panel of Fig.~\ref{fig:experimentalCF} represent the correlation function evaluated from LO \chEFT~\cite{Haidenbauer:2009qn} for a regulator function cut-off of 550 and 700 MeV, respectively. The genuine \LXi\ correlation function is evaluated by using the LL model with the scattering parameters for
$\Lambda$--$\Xi$ provided in Ref.~\cite{Haidenbauer:2009qn}.
The scattering length in this case indicates a rather strong attraction in the singlet channel and a mild repulsion in the triplet channel. The predicted values depend strongly on the cut-off choice, which is reflected in the correlation function. The curve corresponding to the LO \chEFT\ potential with $550$ MeV cut-off, with rather large scattering length, is not compatible with the experimental correlation function. On the other hand, the result for the potential with cut-off $700$ MeV is close to the data. However, this interaction implies the presence of a shallow $\Lambda \Xi$ bound state with a binding energy of just $0.43$ MeV. Such bound states are not seen anymore in the extrapolation of the NLO interactions from $S=-2$~\cite{Haidenbauer:2015zqb,Haidenbauer:2018gvg} to $S=-3$~\cite{Haidenbauer:2022suy}, where effects from SU(3) symmetry breaking have been properly accounted for, in line with the power counting. They are also not supported by the available lattice QCD simulations close to the physical point~\cite{Ishii:2018ddn}.
The correlation functions expected from the NLO calculations, based on NLO16~\cite{Haidenbauer:2015zqb} and NLO19~\cite{Haidenbauer:2018gvg}, are represented by the magenta and green bands in the left panel of Fig.~\ref{fig:experimentalCF}, respectively. The dark and light bands represent the interactions with regulator function cut-offs of 500 and 650 MeV, respectively, for both potentials.
The correlation function was evaluated by using the LL model with the scattering lengths\footnote{For the triplet state in NLO16 the effective range is set to $d_{0} =$ 0 fm. This is necessary since the LL does not provide stable results for the large effective ranges predicted by the theory (see Table~\ref{tab:scatpar}) in combination with the small source radius.
} provided in Ref.~\cite{Haidenbauer:2022suy}.
The NLO19 potential is more attractive than the NLO16 in the triplet channel, and there is also a sizable cut-off dependence. This is reflected in a larger correlation function, in particular for the 500 MeV cut-off (dark green band), that clearly exceeds the data.
This demonstrates that the ALICE data delivers important constraints in the $S=-3$ sector needed to fix the free parameters in the \chEFT\ NLO calculations~\cite{Haidenbauer:2022suy}.

The Nijmegen meson exchange model~\cite{Stoks:1999bz} predicts the existence of a $\Sigma \Xi$ bound state, though in the case of the \LXi\ channel, for 5 of the 6 different versions of the model, a mild attraction and a mild repulsion in the singlet and triplet configurations, respectively, are predicted. The light orange band in Fig.~\ref{fig:experimentalCF} shows the expectation from the version NSC97a. The substantially smaller scattering lengths in the singlet state compared to the \chEFT\ potentials are reflected in a suppressed correlation function that agrees with the data. 

It is worth mentioning that the scattering parameters from the quark constituent model \textit{fss2}~\cite{Fujiwara:2006yh} coincide qualitatively with the Nijmegen potential, although that model does not predict any $\Sigma \Xi$ or $\Lambda \Xi$ bound states.

The right panel of Fig.~\ref{fig:experimentalCF} shows the comparison with the results for the \HALQCD\ \LXi\ potential~\cite{Ishii:2018ddn}. The width of the \HALQCD\ curves include statistical and systematic variations of the lattice calculations following the recipe in Ref.~\cite{ALICE:2020mfd}.
The \HALQCD\ potential is a $\Lambda\Xi$--$\Sigma\Xi$ coupled-channel potential. It presents attraction in both the singlet and the triplet configurations but does not predict the formation of any bound state. The orange curve shows the results from an effective \LXi\ potential, where the coupling to $\Sigma$--$\Xi$ from the lattice simulations is incorporated effectively into the strength  of the \LXi\ interaction.
For reference, the red curve shows the correlation function
from the $\Lambda\Xi$--$\Lambda\Xi$ elastic potential alone, free of effects from the coupling.
The difference between the orange and red curves in Figs.~\ref{fig:experimentalCF} and~\ref{fig:genuine} demonstrates a rather strong coupling. This is particularly noticeable if the results are compared to the \HALQCD\ p$\Xi$--$\Lambda\Lambda$ potential
in the $|S|=2$ sector
~\cite{Sasaki:2019qnh}, in which case the coupling between channels is small and has negligible effects in the correlation function~\cite{Kamiya:2021haz}. While the ALICE data shows better compatibility with the single channel $\Lambda\Xi$--$\Lambda\Xi$ elastic potential, it is not sensitive to the effects of the coupling as shown by Lattice QCD.

Threshold cusp-like structures at the channel opening created by the inelastic channels could be formed~\cite{Haidenbauer:2018jvl} with an amplitude depending on the properties of the interaction, the strength of the coupling between channels, and the amount of initial state pairs in the inelastic channels. The ALICE data do not present significant structures at the kinematic opening of the $\Sigma^-$--$\Xi^0$, $\Sigma^0$--$\Xi^-$ and n--$\Omega^-$ channels, at \kstar~ values of ~303, 308 and 468~\MeVc\, respectively. 

The scattering parameters of all considered interactions are summarized in Table~\ref{tab:scatpar}, together with the compatibility of each of them with the \LXi\ correlation function in terms of
$n_\sigma$
computed in the range \kstar$<200$~\MeVc, 
considering the statistical and systematic uncertainties of the data.
The scattering parameters of the \HALQCD\ calculations were extracted using CATS by fitting the phase shifts $\delta_{0}$ with the effective range approximation~\cite{sakurai_napolitano_2017,griffiths2005introduction}.

\begin{table}[]
\caption{Summary of scattering parameters and agreement with the data in terms of $n_\sigma$ of the considered interactions. The effective range parameters for the Lattice QCD results are obtained from the low-energy phase shifts which are extracted with CATS in the evaluation of the potentials. The agreement of the baseline (no interaction assumed) is also given.
    }
    \centering
\begin{tabular}{|c|c|cc|cc|c|}
\hline
\multirow{2}{*}{potential} & \multirow{2}{*}{cut-off (MeV) / version} & \multicolumn{2}{c|}{\textbf{singlet}} & \multicolumn{2}{c|}{\textbf{triplet}} & \multirow{2}{*}{$n_\sigma$} \\
 &  & \multicolumn{1}{c}{$f_0^0$} & $d_0^0$ & \multicolumn{1}{c}{$f_0^1$} & $d_0^1$ &  \\ \hline
\multirow{2}{*}{$\chi \text{EFT LO}$~\cite{Haidenbauer:2009qn}} & 550 & \multicolumn{1}{c|}{$33.5$} & $1.00$ & \multicolumn{1}{c|}{$-0.33$} & $-0.36$ & $3.06-5.12$ \\
 & 700 & \multicolumn{1}{c|}{$-9.07$} & $0.87$ & \multicolumn{1}{c|}{$-0.31$} & $-0.27$ & $0.78-1.60$ \\ \hline
\multirow{2}{*}{$\chi \text{EFT NLO16}$~\cite{Haidenbauer:2015zqb}} & 500 & \multicolumn{1}{c|}{$0.99$} & $5.77$ & \multicolumn{1}{c|}{$-0.026$} & $142.9$ & $0.56-0.93$ \\
 & 650 & \multicolumn{1}{c|}{$0.91$} & $4.63$ & \multicolumn{1}{c|}{$0.12$} & $32.02$ & $0.91-1.61$ \\ \hline
\multirow{2}{*}{$\chi \text{EFT NLO19}$~\cite{Haidenbauer:2018gvg}} & 500 & \multicolumn{1}{c|}{$0.99$} & $5.77$ & \multicolumn{1}{c|}{$1.66$} & $1.49$ & $5.47-7.26$ \\
 & 650 & \multicolumn{1}{c|}{$0.91$} & $4.63$ & \multicolumn{1}{c|}{$0.42$} & $6.33$ & $1.30-2.10$ \\ \hline
NSC97a~\cite{Stoks:1999bz} &  & \multicolumn{1}{c|}{$0.80$} & $4.71$ & \multicolumn{1}{c|}{$-0.54$} & $-0.47$ & $0.68-1.04$ \\ \hline
\multirow{2}{*}{HAL QCD~\cite{Ishii:2018ddn}} & $\Lambda\Xi$--$\Sigma\Xi$ eff.  & \multicolumn{1}{c|}{$0.60$} & $6.01$ & \multicolumn{1}{c|}{$0.50$} & $5.36$ & $1.43-2.34$ \\
 & $\Lambda\Xi$--$\Lambda\Xi$ only & \multicolumn{1}{c|}{--} & -- & \multicolumn{1}{c|}{--} & -- & $0.64-1.04$ \\ \hline
Baseline &  & \multicolumn{1}{c|}{--} & -- & \multicolumn{1}{c|}{--} & -- & $0.78$ \\ \hline
\end{tabular}    
    \label{tab:scatpar}
\end{table}

\section*{Summary}

This Letter presents the first measurement of the \LXi\ interaction, quantified via the correlation function $C({k^*})$ in high-multiplicity \pp\ collisions at \onethree.
The measured $C({k^*})$ were compared with different descriptions of the \LXi\ interaction, including leading-order and next-to-leading-order chiral Effective Field Theory calculations, a meson exchange model, and recent Lattice QCD calculations.
Despite the limited statistical significance and the contamination from feed-down contributions, the data provide the first constraint for theoretical investigations and are seen to be more compatible with predictions of small scattering parameters and hence a weak \LXi\ interaction.
The limitations of the data sample prevent from drawing further conclusions on the influence of coupled channels in the correlation function, and no significant cusp-like structures are observed at the opening of the $\Sigma^-$--$\Xi^0$, $\Sigma^0$--$\Xi^-$ or n--$\Omega^-$  channels.
The presented data demonstrate that the characteristics of the strong interaction in the $|S|=3$ sector can be investigated with the femtoscopy technique. New measurements with the upgraded ALICE apparatus~\cite{Abelevetal:2014cna} will exploit the data collected during upcoming LHC Run~3 and Run~4~\cite{YellowReport_better,ALICE-PUBLIC-2020-005} and should deliver a precise insight into the \LXi\ interaction, providing valuable information for the search of di-baryon states with strangeness content.



\newenvironment{acknowledgement}{\relax}{\relax}
\begin{acknowledgement}
\section*{Acknowledgements}

The ALICE Collaboration would like to thank all its engineers and technicians for their invaluable contributions to the construction of the experiment and the CERN accelerator teams for the outstanding performance of the LHC complex.
The ALICE Collaboration gratefully acknowledges the resources and support provided by all Grid centres and the Worldwide LHC Computing Grid (WLCG) collaboration.
The ALICE Collaboration acknowledges the following funding agencies for their support in building and running the ALICE detector:
A. I. Alikhanyan National Science Laboratory (Yerevan Physics Institute) Foundation (ANSL), State Committee of Science and World Federation of Scientists (WFS), Armenia;
Austrian Academy of Sciences, Austrian Science Fund (FWF): [M 2467-N36] and Nationalstiftung f\"{u}r Forschung, Technologie und Entwicklung, Austria;
Ministry of Communications and High Technologies, National Nuclear Research Center, Azerbaijan;
Conselho Nacional de Desenvolvimento Cient\'{\i}fico e Tecnol\'{o}gico (CNPq), Financiadora de Estudos e Projetos (Finep), Funda\c{c}\~{a}o de Amparo \`{a} Pesquisa do Estado de S\~{a}o Paulo (FAPESP) and Universidade Federal do Rio Grande do Sul (UFRGS), Brazil;
Bulgarian Ministry of Education and Science, within the National Roadmap for Research Infrastructures 2020-2027 (object CERN), Bulgaria;
Ministry of Education of China (MOEC) , Ministry of Science \& Technology of China (MSTC) and National Natural Science Foundation of China (NSFC), China;
Ministry of Science and Education and Croatian Science Foundation, Croatia;
Centro de Aplicaciones Tecnol\'{o}gicas y Desarrollo Nuclear (CEADEN), Cubaenerg\'{\i}a, Cuba;
Ministry of Education, Youth and Sports of the Czech Republic, Czech Republic;
The Danish Council for Independent Research | Natural Sciences, the VILLUM FONDEN and Danish National Research Foundation (DNRF), Denmark;
Helsinki Institute of Physics (HIP), Finland;
Commissariat \`{a} l'Energie Atomique (CEA) and Institut National de Physique Nucl\'{e}aire et de Physique des Particules (IN2P3) and Centre National de la Recherche Scientifique (CNRS), France;
Bundesministerium f\"{u}r Bildung und Forschung (BMBF) and GSI Helmholtzzentrum f\"{u}r Schwerionenforschung GmbH, Germany;
General Secretariat for Research and Technology, Ministry of Education, Research and Religions, Greece;
National Research, Development and Innovation Office, Hungary;
Department of Atomic Energy Government of India (DAE), Department of Science and Technology, Government of India (DST), University Grants Commission, Government of India (UGC) and Council of Scientific and Industrial Research (CSIR), India;
National Research and Innovation Agency - BRIN, Indonesia;
Istituto Nazionale di Fisica Nucleare (INFN), Italy;
Japanese Ministry of Education, Culture, Sports, Science and Technology (MEXT) and Japan Society for the Promotion of Science (JSPS) KAKENHI, Japan;
Consejo Nacional de Ciencia (CONACYT) y Tecnolog\'{i}a, through Fondo de Cooperaci\'{o}n Internacional en Ciencia y Tecnolog\'{i}a (FONCICYT) and Direcci\'{o}n General de Asuntos del Personal Academico (DGAPA), Mexico;
Nederlandse Organisatie voor Wetenschappelijk Onderzoek (NWO), Netherlands;
The Research Council of Norway, Norway;
Commission on Science and Technology for Sustainable Development in the South (COMSATS), Pakistan;
Pontificia Universidad Cat\'{o}lica del Per\'{u}, Peru;
Ministry of Education and Science, National Science Centre and WUT ID-UB, Poland;
Korea Institute of Science and Technology Information and National Research Foundation of Korea (NRF), Republic of Korea;
Ministry of Education and Scientific Research, Institute of Atomic Physics, Ministry of Research and Innovation and Institute of Atomic Physics and University Politehnica of Bucharest, Romania;
Ministry of Education, Science, Research and Sport of the Slovak Republic, Slovakia;
National Research Foundation of South Africa, South Africa;
Swedish Research Council (VR) and Knut \& Alice Wallenberg Foundation (KAW), Sweden;
European Organization for Nuclear Research, Switzerland;
Suranaree University of Technology (SUT), National Science and Technology Development Agency (NSTDA), Thailand Science Research and Innovation (TSRI) and National Science, Research and Innovation Fund (NSRF), Thailand;
Turkish Energy, Nuclear and Mineral Research Agency (TENMAK), Turkey;
National Academy of  Sciences of Ukraine, Ukraine;
Science and Technology Facilities Council (STFC), United Kingdom;
National Science Foundation of the United States of America (NSF) and United States Department of Energy, Office of Nuclear Physics (DOE NP), United States of America.
In addition, individual groups or members have received support from:
Marie Sk\l{}odowska Curie, Strong 2020 - Horizon 2020, European Research Council (grant nos. 824093, 896850, 950692), European Union;
Academy of Finland (Center of Excellence in Quark Matter) (grant nos. 346327, 346328), Finland;
Programa de Apoyos para la Superaci\'{o}n del Personal Acad\'{e}mico, UNAM, Mexico.
\end{acknowledgement}

\bibliographystyle{utphys}   
\bibliography{Content/Bibliography.bib}

\newpage
\appendix

\section{The ALICE Collaboration}
\label{app:collab}
\begin{flushleft} 
\small

S.~Acharya\,\orcidlink{0000-0002-9213-5329}\,$^{\rm 127,134}$, 
D.~Adamov\'{a}\,\orcidlink{0000-0002-0504-7428}\,$^{\rm 87}$, 
A.~Adler$^{\rm 70}$, 
G.~Aglieri Rinella\,\orcidlink{0000-0002-9611-3696}\,$^{\rm 32}$, 
M.~Agnello\,\orcidlink{0000-0002-0760-5075}\,$^{\rm 29}$, 
N.~Agrawal\,\orcidlink{0000-0003-0348-9836}\,$^{\rm 50}$, 
Z.~Ahammed\,\orcidlink{0000-0001-5241-7412}\,$^{\rm 134}$, 
S.~Ahmad\,\orcidlink{0000-0003-0497-5705}\,$^{\rm 15}$, 
S.U.~Ahn\,\orcidlink{0000-0001-8847-489X}\,$^{\rm 71}$, 
I.~Ahuja\,\orcidlink{0000-0002-4417-1392}\,$^{\rm 37}$
% ...

\section*{Affiliation Notes}

$^{\rm I}$ Deceased\\
$^{\rm II}$ Also at: Max-Planck-Institut f\"{u}r Physik, Munich, Germany\\
$^{\rm III}$ Also at: Italian National Agency for New Technologies, Energy and Sustainable Economic Development (ENEA), Bologna, Italy\\
$^{\rm IV}$ Also at: Dipartimento DET del Politecnico di Torino, Turin, Italy\\
$^{\rm V}$ Also at: Department of Applied Physics, Aligarh Muslim University, Aligarh, India\\
$^{\rm VI}$ Also at: Institute of Theoretical Physics, University of Wroclaw, Poland\\
$^{\rm VII}$ Also at: An institution covered by a cooperation agreement with CERN\\

\section*{Collaboration Institutes}

$^{1}$ A.I. Alikhanyan National Science Laboratory (Yerevan Physics Institute) Foundation, Yerevan, Armenia\\
$^{2}$ AGH University of Science and Technology, Cracow, Poland\\
$^{3}$ Bogolyubov Institute for Theoretical Physics, National Academy of Sciences of Ukraine, Kiev, Ukraine\\
$^{4}$ Bose Institute, Department of Physics  and Centre for Astroparticle Physics and Space Science (CAPSS), Kolkata, India\\
$^{5}$ California Polytechnic State University, San Luis Obispo, California, United States\\
$^{6}$ Central China Normal University, Wuhan, China\\
$^{7}$ Centro de Aplicaciones Tecnol\'{o}gicas y Desarrollo Nuclear (CEADEN), Havana, Cuba\\
$^{8}$ Centro de Investigaci\'{o}n y de Estudios Avanzados (CINVESTAV), Mexico City and M\'{e}rida, Mexico\\
$^{9}$ Chicago State University, Chicago, Illinois, United States\\
$^{10}$ China Institute of Atomic Energy, Beijing, China\\
$^{11}$ Chungbuk National University, Cheongju, Republic of Korea\\
$^{12}$ Comenius University Bratislava, Faculty of Mathematics, Physics and Informatics, Bratislava, Slovak Republic\\
$^{13}$ COMSATS University Islamabad, Islamabad, Pakistan\\
$^{14}$ Creighton University, Omaha, Nebraska, United States\\
$^{15}$ Department of Physics, Aligarh Muslim University, Aligarh, India\\
$^{16}$ Department of Physics, Pusan National University, Pusan, Republic of Korea\\
$^{17}$ Department of Physics, Sejong University, Seoul, Republic of Korea\\
$^{18}$ Department of Physics, University of California, Berkeley, California, United States\\
$^{19}$ Department of Physics, University of Oslo, Oslo, Norway\\
$^{20}$ Department of Physics and Technology, University of Bergen, Bergen, Norway\\
$^{21}$ Dipartimento di Fisica, Universit\`{a} di Pavia, Pavia, Italy\\
$^{22}$ Dipartimento di Fisica dell'Universit\`{a} and Sezione INFN, Cagliari, Italy\\
$^{23}$ Dipartimento di Fisica dell'Universit\`{a} and Sezione INFN, Trieste, Italy\\
$^{24}$ Dipartimento di Fisica dell'Universit\`{a} and Sezione INFN, Turin, Italy\\
$^{25}$ Dipartimento di Fisica e Astronomia dell'Universit\`{a} and Sezione INFN, Bologna, Italy\\
$^{26}$ Dipartimento di Fisica e Astronomia dell'Universit\`{a} and Sezione INFN, Catania, Italy\\
$^{27}$ Dipartimento di Fisica e Astronomia dell'Universit\`{a} and Sezione INFN, Padova, Italy\\
$^{28}$ Dipartimento di Fisica `E.R.~Caianiello' dell'Universit\`{a} and Gruppo Collegato INFN, Salerno, Italy\\
$^{29}$ Dipartimento DISAT del Politecnico and Sezione INFN, Turin, Italy\\
$^{30}$ Dipartimento di Scienze MIFT, Universit\`{a} di Messina, Messina, Italy\\
$^{31}$ Dipartimento Interateneo di Fisica `M.~Merlin' and Sezione INFN, Bari, Italy\\
$^{32}$ European Organization for Nuclear Research (CERN), Geneva, Switzerland\\
$^{33}$ Faculty of Electrical Engineering, Mechanical Engineering and Naval Architecture, University of Split, Split, Croatia\\
$^{34}$ Faculty of Engineering and Science, Western Norway University of Applied Sciences, Bergen, Norway\\
$^{35}$ Faculty of Nuclear Sciences and Physical Engineering, Czech Technical University in Prague, Prague, Czech Republic\\
$^{36}$ Faculty of Physics, Sofia University, Sofia, Bulgaria\\
$^{37}$ Faculty of Science, P.J.~\v{S}af\'{a}rik University, Ko\v{s}ice, Slovak Republic\\
$^{38}$ Frankfurt Institute for Advanced Studies, Johann Wolfgang Goethe-Universit\"{a}t Frankfurt, Frankfurt, Germany\\
$^{39}$ Fudan University, Shanghai, China\\
$^{40}$ Gangneung-Wonju National University, Gangneung, Republic of Korea\\
$^{41}$ Gauhati University, Department of Physics, Guwahati, India\\
$^{42}$ Helmholtz-Institut f\"{u}r Strahlen- und Kernphysik, Rheinische Friedrich-Wilhelms-Universit\"{a}t Bonn, Bonn, Germany\\
$^{43}$ Helsinki Institute of Physics (HIP), Helsinki, Finland\\
$^{44}$ High Energy Physics Group,  Universidad Aut\'{o}noma de Puebla, Puebla, Mexico\\
$^{45}$ Horia Hulubei National Institute of Physics and Nuclear Engineering, Bucharest, Romania\\
$^{46}$ Indian Institute of Technology Bombay (IIT), Mumbai, India\\
$^{47}$ Indian Institute of Technology Indore, Indore, India\\
$^{48}$ INFN, Laboratori Nazionali di Frascati, Frascati, Italy\\
$^{49}$ INFN, Sezione di Bari, Bari, Italy\\
$^{50}$ INFN, Sezione di Bologna, Bologna, Italy\\
$^{51}$ INFN, Sezione di Cagliari, Cagliari, Italy\\
$^{52}$ INFN, Sezione di Catania, Catania, Italy\\
$^{53}$ INFN, Sezione di Padova, Padova, Italy\\
$^{54}$ INFN, Sezione di Pavia, Pavia, Italy\\
$^{55}$ INFN, Sezione di Torino, Turin, Italy\\
$^{56}$ INFN, Sezione di Trieste, Trieste, Italy\\
$^{57}$ Inha University, Incheon, Republic of Korea\\
$^{58}$ Institute for Advanced Simulation, Forschungszentrum J\"ulich, J\"ulich, Germany\\
$^{59}$ Institute for Gravitational and Subatomic Physics (GRASP), Utrecht University/Nikhef, Utrecht, Netherlands\\
$^{60}$ Institute of Experimental Physics, Slovak Academy of Sciences, Ko\v{s}ice, Slovak Republic\\
$^{61}$ Institute of Physics, Homi Bhabha National Institute, Bhubaneswar, India\\
$^{62}$ Institute of Physics of the Czech Academy of Sciences, Prague, Czech Republic\\
$^{63}$ Institute of Space Science (ISS), Bucharest, Romania\\
$^{64}$ Institut f\"{u}r Kernphysik, Johann Wolfgang Goethe-Universit\"{a}t Frankfurt, Frankfurt, Germany\\
$^{65}$ Instituto de Ciencias Nucleares, Universidad Nacional Aut\'{o}noma de M\'{e}xico, Mexico City, Mexico\\
$^{66}$ Instituto de F\'{i}sica, Universidade Federal do Rio Grande do Sul (UFRGS), Porto Alegre, Brazil\\
$^{67}$ Instituto de F\'{\i}sica, Universidad Nacional Aut\'{o}noma de M\'{e}xico, Mexico City, Mexico\\
$^{68}$ iThemba LABS, National Research Foundation, Somerset West, South Africa\\
$^{69}$ Jeonbuk National University, Jeonju, Republic of Korea\\
$^{70}$ Johann-Wolfgang-Goethe Universit\"{a}t Frankfurt Institut f\"{u}r Informatik, Fachbereich Informatik und Mathematik, Frankfurt, Germany\\
$^{71}$ Korea Institute of Science and Technology Information, Daejeon, Republic of Korea\\
$^{72}$ KTO Karatay University, Konya, Turkey\\
$^{73}$ Laboratoire de Physique des 2 Infinis, Ir\`{e}ne Joliot-Curie, Orsay, France\\
$^{74}$ Laboratoire de Physique Subatomique et de Cosmologie, Universit\'{e} Grenoble-Alpes, CNRS-IN2P3, Grenoble, France\\
$^{75}$ Lawrence Berkeley National Laboratory, Berkeley, California, United States\\
$^{76}$ Lund University Department of Physics, Division of Particle Physics, Lund, Sweden\\
$^{77}$ Nagasaki Institute of Applied Science, Nagasaki, Japan\\
$^{78}$ Nara Women{'}s University (NWU), Nara, Japan\\
$^{79}$ National and Kapodistrian University of Athens, School of Science, Department of Physics , Athens, Greece\\
$^{80}$ National Centre for Nuclear Research, Warsaw, Poland\\
$^{81}$ National Institute of Science Education and Research, Homi Bhabha National Institute, Jatni, India\\
$^{82}$ National Nuclear Research Center, Baku, Azerbaijan\\
$^{83}$ National Research and Innovation Agency - BRIN, Jakarta, Indonesia\\
$^{84}$ Niels Bohr Institute, University of Copenhagen, Copenhagen, Denmark\\
$^{85}$ Nikhef, National institute for subatomic physics, Amsterdam, Netherlands\\
$^{86}$ Nuclear Physics Group, STFC Daresbury Laboratory, Daresbury, United Kingdom\\
$^{87}$ Nuclear Physics Institute of the Czech Academy of Sciences, Husinec-\v{R}e\v{z}, Czech Republic\\
$^{88}$ Oak Ridge National Laboratory, Oak Ridge, Tennessee, United States\\
$^{89}$ Ohio State University, Columbus, Ohio, United States\\
$^{90}$ Physics department, Faculty of science, University of Zagreb, Zagreb, Croatia\\
$^{91}$ Physics Department, Panjab University, Chandigarh, India\\
$^{92}$ Physics Department, University of Jammu, Jammu, India\\
$^{93}$ Physics Department, University of Rajasthan, Jaipur, India\\
$^{94}$ Physics Program and International Institute for Sustainability with Knotted Chiral Meta Matter (SKCM2), Hiroshima University, Hiroshima, Japan\\
$^{95}$ Physikalisches Institut, Eberhard-Karls-Universit\"{a}t T\"{u}bingen, T\"{u}bingen, Germany\\
$^{96}$ Physikalisches Institut, Ruprecht-Karls-Universit\"{a}t Heidelberg, Heidelberg, Germany\\
$^{97}$ Physik Department, Technische Universit\"{a}t M\"{u}nchen, Munich, Germany\\
$^{98}$ Politecnico di Bari and Sezione INFN, Bari, Italy\\
$^{99}$ Research Center for Nuclear Physics, Osaka University, Osaka, Japan\\
$^{100}$ Research Division and ExtreMe Matter Institute EMMI, GSI Helmholtzzentrum f\"ur Schwerionenforschung GmbH, Darmstadt, Germany\\
$^{101}$ RIKEN iTHEMS, Wako, Japan\\
$^{102}$ Saha Institute of Nuclear Physics, Homi Bhabha National Institute, Kolkata, India\\
$^{103}$ School of Physics and Astronomy, University of Birmingham, Birmingham, United Kingdom\\
$^{104}$ Secci\'{o}n F\'{\i}sica, Departamento de Ciencias, Pontificia Universidad Cat\'{o}lica del Per\'{u}, Lima, Peru\\
$^{105}$ Stefan Meyer Institut f\"{u}r Subatomare Physik (SMI), Vienna, Austria\\
$^{106}$ SUBATECH, IMT Atlantique, Nantes Universit\'{e}, CNRS-IN2P3, Nantes, France\\
$^{107}$ Suranaree University of Technology, Nakhon Ratchasima, Thailand\\
$^{108}$ Technical University of Ko\v{s}ice, Ko\v{s}ice, Slovak Republic\\
$^{109}$ The Henryk Niewodniczanski Institute of Nuclear Physics, Polish Academy of Sciences, Cracow, Poland\\
$^{110}$ The University of Texas at Austin, Austin, Texas, United States\\
$^{111}$ Universidad Aut\'{o}noma de Sinaloa, Culiac\'{a}n, Mexico\\
$^{112}$ Universidade de S\~{a}o Paulo (USP), S\~{a}o Paulo, Brazil\\
$^{113}$ Universidade Estadual de Campinas (UNICAMP), Campinas, Brazil\\
$^{114}$ Universidade Federal do ABC, Santo Andre, Brazil\\
$^{115}$ University of Cape Town, Cape Town, South Africa\\
$^{116}$ University of Houston, Houston, Texas, United States\\
$^{117}$ University of Jyv\"{a}skyl\"{a}, Jyv\"{a}skyl\"{a}, Finland\\
$^{118}$ University of Kansas, Lawrence, Kansas, United States\\
$^{119}$ University of Liverpool, Liverpool, United Kingdom\\
$^{120}$ University of Science and Technology of China, Hefei, China\\
$^{121}$ University of South-Eastern Norway, Kongsberg, Norway\\
$^{122}$ University of Tennessee, Knoxville, Tennessee, United States\\
$^{123}$ University of the Witwatersrand, Johannesburg, South Africa\\
$^{124}$ University of Tokyo, Tokyo, Japan\\
$^{125}$ University of Tsukuba, Tsukuba, Japan\\
$^{126}$ University Politehnica of Bucharest, Bucharest, Romania\\
$^{127}$ Universit\'{e} Clermont Auvergne, CNRS/IN2P3, LPC, Clermont-Ferrand, France\\
$^{128}$ Universit\'{e} de Lyon, CNRS/IN2P3, Institut de Physique des 2 Infinis de Lyon, Lyon, France\\
$^{129}$ Universit\'{e} de Strasbourg, CNRS, IPHC UMR 7178, F-67000 Strasbourg, France, Strasbourg, France\\
$^{130}$ Universit\'{e} Paris-Saclay Centre d'Etudes de Saclay (CEA), IRFU, D\'{e}partment de Physique Nucl\'{e}aire (DPhN), Saclay, France\\
$^{131}$ Universit\`{a} degli Studi di Foggia, Foggia, Italy\\
$^{132}$ Universit\`{a} del Piemonte Orientale, Vercelli, Italy\\
$^{133}$ Universit\`{a} di Brescia, Brescia, Italy\\
$^{134}$ Variable Energy Cyclotron Centre, Homi Bhabha National Institute, Kolkata, India\\
$^{135}$ Warsaw University of Technology, Warsaw, Poland\\
$^{136}$ Wayne State University, Detroit, Michigan, United States\\
$^{137}$ Westf\"{a}lische Wilhelms-Universit\"{a}t M\"{u}nster, Institut f\"{u}r Kernphysik, M\"{u}nster, Germany\\
$^{138}$ Wigner Research Centre for Physics, Budapest, Hungary\\
$^{139}$ Yale University, New Haven, Connecticut, United States\\
$^{140}$ Yonsei University, Seoul, Republic of Korea\\
$^{141}$  Zentrum  f\"{u}r Technologie und Transfer (ZTT), Worms, Germany\\
$^{142}$ Affiliated with an institute covered by a cooperation agreement with CERN\\
$^{143}$ Affiliated with an international laboratory covered by a cooperation agreement with CERN.\\

\end{flushleft} 
\end{document}